\documentclass{PoS}
\input{babarsym}
\newcommand{\kpi}{$K\pi^-$}

\newcommand{\psipi}{$\psi\pi^-$}
\newcommand{\jpsipi}{$J/\psi\pi^-$}
\newcommand{\psitwospi}{$\psi(2S)\pi^-$}

\newcommand{\Ksone}{$K^{\ast}(892)$}
\newcommand{\Kstwo}{$K^{\ast}_2(1430)$}

\newcommand{\z}{$Z(4430)^-$}
\newcommand{\costhk}{$\cos\theta_K$}

\title{Studies of charmonium-like states at \babar\/}

\ShortTitle{Studies of charmonium-like states at \babar\/}

\author{\speaker{Arafat Gabareen Mokhtar}\thanks{On behalf of the \babar\ Collaboration}\\
        SLAC National Accelerator Laboratory\\
	2575 Sand Hill Rd, Menlo Park, CA 94025, USA\\
        E-mail: \email{mokhtar@slac.stanford.edu}}

\abstract{Several charmonium-like states above $D\bar{D}$ threshold
          have been discovered at the Belle and \babar\
          $B$-factories. Some of these states are produced via Initial
          State Radiation ({\it e.g.} $Y(4260)$ and $Y(4350)$), and
          some are observed in $B$-meson decays ({\it e.g.}
          $X(3872)$, and $Y(3940)$). The Belle observations of the
          enhancements in the $\psi(2S)\pi^-$ and $\chi_{c1}\pi^-$,
          {\it i. e.} the $Z(4430)^-$, $Z_1(4050)^-$, and
          $Z_2(4250)^-$, have generated a great deal of interest,
          because such states must have minimum quark content
          $(c\bar{c}d\bar{u})$, {\it i.e.} these are four-quark
          states. The \babar\ Collaboration does not confirm the
          existinence of the \z\/.}

\FullConference{The 2009 Europhysics Conference on High Energy Physics,\\
		 July 16 - 22 2009\\
		 Krakow, Poland}

\begin{document}

\section{Introduction}
\begin{figure}[!htbp]
  \begin{center}
    \includegraphics[width=10.cm]{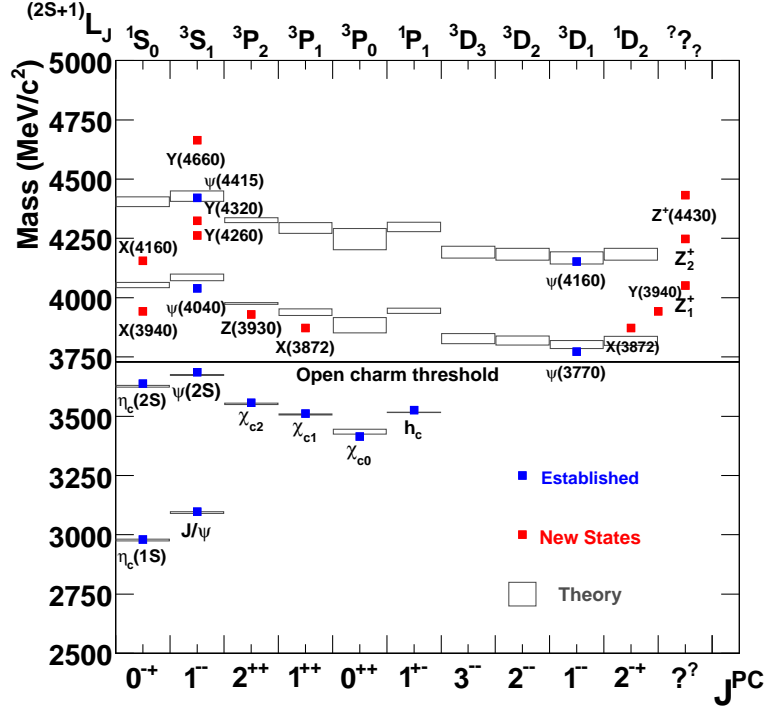}
    \caption{The mass versus the quantum numbers ($J^{PC}$) for the
    charmonium-like states. The boxes represent the predictions; blue
    boxes show the established states, and the red boxes indicate the
    new states discovered at the $B$-factories.}
    \label{fig:spec}
\end{center}
\end{figure}

Many charmonium states have been predicted by a variety of theoretical
models. All predicted states below open-charm threshold have been
discovered. At the $B$-factories, many new states above open-charm
threshold have been discovered, but no new candidate states have been
reported below this threshold. An overview of the charmonium and
charmonium-like states is shown in Fig.~\ref{fig:spec}. The quantum
numbers ($J^{PC}$) of some of the new states have not yet been
determined.

\section{The $Y(4260)\rightarrow J/\psi\pi^+\pi^-$}
The $Y(4260)$ was discovered~\cite{Aubert:2005rm} by \babar\ in the
Initial State Radiation (ISR) process $e^+e^-\rightarrow
\gamma_{ISR}Y(4260)$, $Y(4260)\rightarrow J/\psi\pi^+\pi^-$. Being
formed directly in $e^+e^-$ annihilation, this should be a
$J^{PC}=1^{--}$ state. However, its nature is not yet understood, and
does not fit into a simple charmonium model. The \babar\ Collaboration
is finalizing an update, using the full dataset with 454
fb$^{-1}$~\cite{:2008ic}, on the decay mode $Y(4260)\rightarrow
J/\psi\pi^+\pi^-$. The preliminary updated $Y(4260)$ mass and width
values are $m=4252\pm6(stat)^{+2}_{-3}(syst)$ \mevcc\/ and
$\Gamma=105\pm18(stat)^{+4}_{-6}(syst)$ \mev\/, respectively. There is
no evidence for the enhancement at $\sim 4005$ \mevcc\ reported by
the Belle Collaboration~\cite{:2007sj}.

\begin{figure}[!htbp]
  \begin{center}
    \includegraphics[width=8.cm]{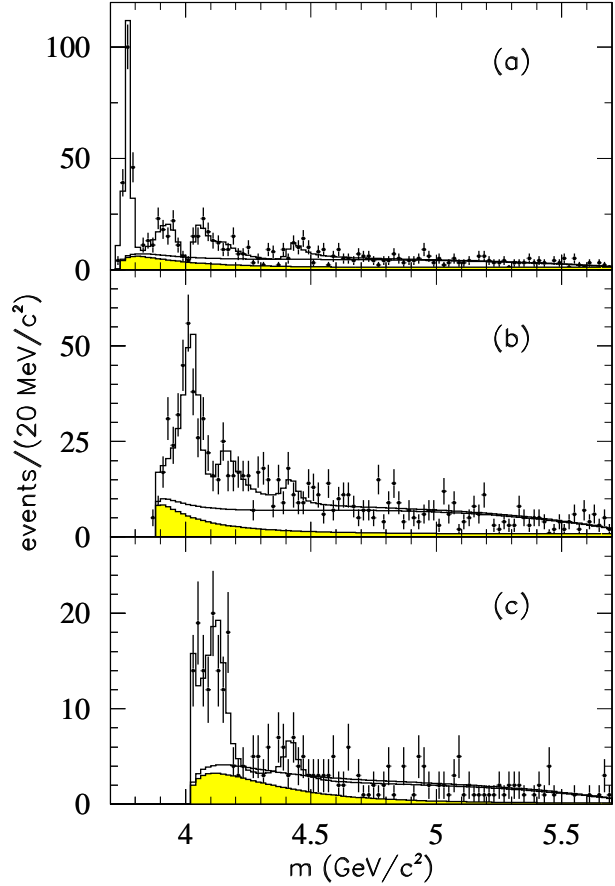}
    \caption{Fits to the (a) $D\bar{D}$, (b) $D^{\ast}\bar{D}$, and
    (c) $D^{\ast}\bar{D^{\ast}}$ mass spectra. The points represent
    the data and the curves show the fitted functions. The shaded
    histograms correspond to the smoothed incoherent background
    estimated from sidebands. The smooth solid curves represent the
    nonresonant contributions.}
    \label{fig:y4260}
\end{center}
\end{figure}

The \babar\ Collaboration has searched also for the decays
$Y(4260)\rightarrow D\bar{D}$, $Y(4260)\rightarrow D^{\ast}\bar{D}$,
and $Y(4260)\rightarrow D^{\ast}\bar{D^{\ast}}$ in ISR
events~\cite{YtoDD}. The $D\bar{D}$ mass distribution for these decay
modes is shown in Fig.~\ref{fig:y4260}. No evidence for the $Y(4260)$
is observed in any of these distributions.

\section{The $Y(4350)\rightarrow \psi(2S)\pi^+\pi^-$}
\babar\ has searched for the decay $Y(4260)\rightarrow
\psi(2S)\pi^+\pi^-$ using a dataset of 298
fb$^{-1}$~\cite{Aubert:2006ge}. A new structure at $\sim 4.32$ \gevcc\
in the $\psi(2S)\pi^+\pi^-$ invariant-mass spectrum, not consistent
with the decay $\psi(4415)\rightarrow \psi(2S)\pi^+\pi^-$, has been
observed. The structure also differs from that reported for the
$Y(4260)$ in Ref.~\cite{Aubert:2005rm}. The possibility that it
represents evidence for a new decay mode for the $Y(4260)$ cannot be
entirely ruled out. The mass and width obtained from a fit to this
structure are $m=4324\pm 24$ \mevcc\ and $\Gamma=172\pm 33$ \mev\/.

\section{The $X(3872)\rightarrow J/\psi\gamma$ and $X(3872)\rightarrow\psi(2S)\gamma$}
The $X(3872)$ was discovered by the Belle Collaboration in the decay
mode $B\rightarrow X(3872)K$, $X(3872)\rightarrow
J/\psi\pi^+\pi^-$~\cite{Choi:2003ue}. Confirmation of this observation
came from CDF, D0, and \babar\
experiments~\cite{Acosta:2003zx,Abazov:2004kp,Aubert:2004ns}. Subsequently
other decay modes have been reported such as $X(3872)\rightarrow
D^0\bar{D^{\ast}}$~\cite{Gokhroo:2006bt,Aubert:2007rva}, and
$X(3872)\rightarrow J/\psi\gamma$~\cite{Abe:2005ix,Aubert:2006aj}. The
\babar\ Collaboration has updated the measurement of
$X(3872)\rightarrow J/\psi\gamma$, and also reported a new decay mode
$X(3872)\rightarrow \psi(2S)\gamma$ using a dataset corresponding to
luminosity of 424 fb$^{-1}$~\cite{:2008rn}. We measure the branching
fractions ${\mathcal{B}}(B^{\pm}\rightarrow X(3872)K^{\pm})\times
\mathcal{B}(X(3872)\rightarrow J/\psi\gamma)=(2.8\pm 0.8(stat)\pm
0.2(syst))\times 10^{-6}$ and $\mathcal{B}(B^{\pm}\rightarrow
X(3872)K^{\pm})\times \mathcal{B}(X(3872)\rightarrow
\psi(2S)\gamma)=(9.9\pm2.9(stat)\pm 0.6(syst))\times 10^{-6}$. The
resulting branching fraction ratio contradicts theoretical
expectation~\cite{Swanson:2006st}. Observation of the decays
$X(3872)\rightarrow J/\psi\gamma$ and $X(3872)\rightarrow
\psi(2S)\gamma$ establishes positive C-parity for the $X(3872)$.

\section{The $Y(3940)\rightarrow J/\psi\omega$}
The Belle Collaboration reported evidence for the $Y(3940)$ in the
decay $B\rightarrow Y(3940)K$, $Y(3940)\rightarrow
J/\psi\omega$~\cite{Abe:2004zs}, with mass and width $3943\pm
11(stat)\pm 13(syst)$ \mevcc\ and $87\pm 22(stat)\pm 26(syst)$ \mev\/,
respectively. The \babar\ Collaboration confirmed~\cite{Aubert:2007vj}
the existence of the $Y(3940)$ using a data sample of 348 fb$^{-1}$,
but measured a lower mass ($3914.6^{+3.8}_{-3.4}(stat)\pm 2(syst)$
\mevcc\/) and smaller width ($34^{+12}_{-8}(stat)\pm 5(syst)$ \mev\/)
than in the Belle analysis. The Belle Collaboration has recently
reported~\cite{OlsenTalk} a new state observed in two-photon
production with a mass of $3914\pm 3(stat)\pm 2(syst)$ \mevcc\/, and a
width of $23\pm 10(stat)^{+2}_{-8}(syst)$ \mev\/. These mass and width
values are in a good agreement with the \babar\ values for the
$Y(3940)$, and so might indicate that these are different decay modes
of the same state.  The \babar\ ratio of the $B^0$ and $B^+$ branching
fractions for the $Y(3940)$ is $0.27^{+0.28}_ {-0.23}(stat)^{+0.04}
_{-0.01}(syst)$. The central value is thus three standard deviations
below the isospin expectation, but agrees well with the corresponding
$X(3872)$ ratio from \babar\/~\cite{Aubert:2008gu}.

\section{Search for the \z\/}
The Belle Collaboration report~\cite{:2007wga} of a charged
charmonium-like structure, the \z\/, in the $\psi(2S)\pi^-$ system
produced in the decays $B^{-,0}\rightarrow\psi(2S)\pi^-K^{0,-}$ has
generated a great deal of interest. If confirmed, such a state must
have minimum quark content ($c\bar{c}d\bar{u}$) so that it would
represent the unequivocal manifestation of a four-quark meson state.

The \babar\/ Collaboration has analyzed a data sample collected at the
$Y(4S)$ resonance (413 fb$^{-1}$) to search for the \z\ state in four
decay modes $B\rightarrow \psi\pi^- K$, where $\psi=J/\psi$ or
$\psi(2S)$ and $K=K^0_S$ or $K^+$. We represent the \kpi\ mass
dependence of the angular structure in the \kpi\ angular distribution
at a given $m_{K\pi^-}$ in terms of Legendre polynomials,
$P_l(\cos\theta_K)$, where the angle $\theta_K$ is between the $K$ in
the \kpi\ rest frame and the \kpi\ direction in the $B$ rest frame. A
backward-forward asymmetry in $\cos\theta_K$ has been measured, and
the details of the observed \kpi\ mass and angular structure have been
carefully taken into account in evaluating their impact on the
corresponding \psipi\ mass distribution.

\begin{figure}[]
  \begin{center}
    \includegraphics[width=13.0cm]{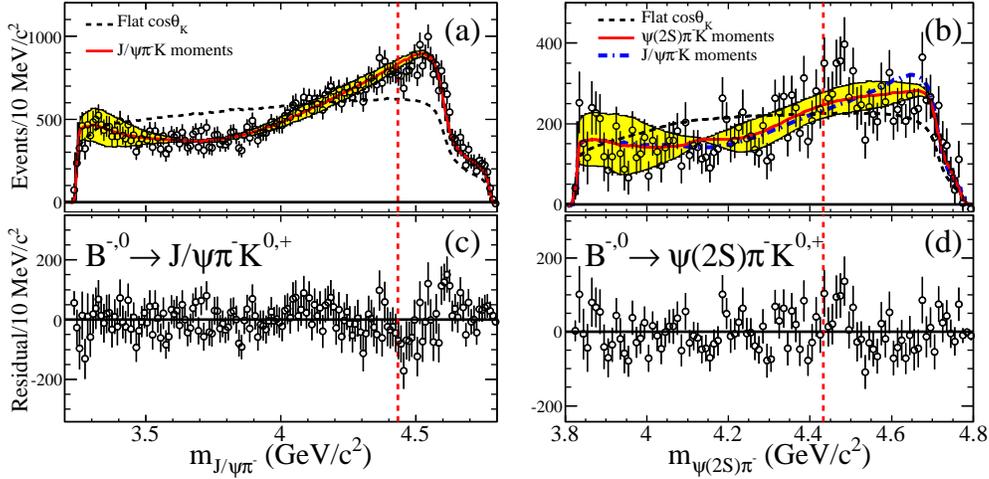}
    \caption{The \psipi\ mass distributions for the combined decay
      modes (a) $B^{-,0}\rightarrow J/\psi\pi^- K^{0,+}$ and (b)
      $B^{-,0}\rightarrow \psi(2S)\pi^- K^{0,+}$. The points show the
      data after efficiency correction and \DeltaE\ sideband
      subtraction. The dashed curves show the \kpi\ reflection for a
      flat \costhk\ distribution, while the solid curves show the
      result of \costhk\ weighting.  The shaded bands represent the
      effect of statistical uncertainty on the normalized moments. In
      (b), the dot-dashed curve indicates the effect of weighting with
      the normalized $J/\psi\pi^-K$ moments. The dashed vertical lines
      indicate the value of $m_{\psi\pi^-}=4.433$ \gevcc\/. In (c) and
      (d), we show the residuals (data-solid curve) for (a) and (b),
      respectively.}
    \label{fig:compare}
\end{center}
\end{figure}

We compare the data and the reflections of the \kpi\ structure into
the \psipi\ mass distributions for the five \kpi\ regions relevant to
the Belle analysis (see Fig. 25 in Ref.~\cite{:2008nk}): below the
\Ksone\/; within 100 \mevcc\ of the \Ksone\/; between the \Ksone\ and
the \Kstwo\/; within 100 \mevcc\ of the \Kstwo\/; and above the
\Kstwo\/. Overall, good agreement between data and \kpi\ reflection is
obtained in the different \kpi\ regions, indicating that no additional
structure is needed to describe the data. In Fig.~\ref{fig:compare},
we show the \psipi\ mass distributions for all of the data. No
compelling evidence for the \z\ is observed.

\begin{figure}[!htbp]
  \begin{center}
    \includegraphics[width=13.cm]{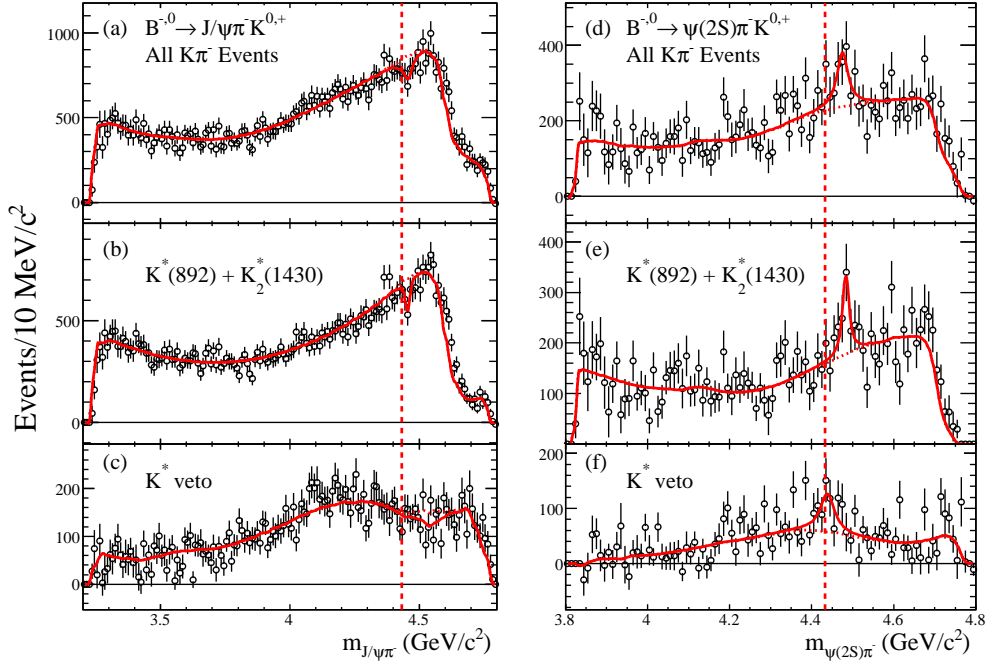}
    \caption{The results of the fits to the corrected mass
      distributions, (a)-(c) for \jpsipi\/, and (d)-(f) for
      \psitwospi\/. The curves are described in the text; the dashed
      vertical lines indicate $m_{\psi\pi^-}=4.433$ \gevcc\/.}
    \label{fig:fits}
\end{center}
\end{figure}

In Fig.~\ref{fig:fits} we show fits to the \psipi\ mass distributions
in which the \kpi\ background shape is fixed and an $S$-wave Breit
Wigner (BW) is used as a signal function. The BW parameters are free
in the fits. Figures~\ref{fig:fits}(a) and (d) are for the entire data
samples; Figures~\ref{fig:fits}(b) and (e) are for the $K^{\ast}$
regions, and Figs~\ref{fig:fits}(c) and (f) are for the \Ksone\ and
\Kstwo\ veto regions combined (the Belle selection). For the \jpsi\
samples, no evidence for any enhancement is obtained. For the
\psitwos\ data small signals are obtained, but their significance is
only in the $2-3\sigma$ range, and in Fig.~\ref{fig:fits}(d) and (e)
the fitted mass is significantly different from the Belle value. In
Fig.~\ref{fig:fits}(f), the signal mass and width are consistent with
the Belle values, but the signal significance is only $1.9\sigma$. We
conclude that the \babar\ data provide no significant evidence for the
existence of the \z\/.

\end{document}